\documentclass[a4paper]{article}
\usepackage[utf8]{inputenc}
\usepackage[T1]{polski}
\usepackage{graphicx}\graphicspath{{img/}}
\usepackage{lmodern,fancyhdr,secdot,float,amsmath,amsfonts,amsthm,amssymb}
\usepackage{booktabs,tabularx}
\usepackage[margin=2cm]{geometry}
\usepackage{multicol}
\usepackage[hidelinks,unicode]{hyperref}
\usepackage{titlesec}
\titleformat{\section}{\centering\normalsize\bf}{\thesection.}{.5em}{\MakeUppercase}
\titleformat*{\subsection}{\bf\normalsize\selectfont}
\titleformat*{\subsubsection}{\bf\normalsize\selectfont}
\newcommand{\titlePL}[1]{\large\textbf{ #1}}
\newcommand{\titleEN}[1]{\normalsize #1}
\newcommand{\keywordsPL}[1]{\small\textbf{Słowa kluczowe:} #1}
\newcommand{\keywordsEN}[1]{\small\textbf{Keywords:} #1}
\newcommand{\abstractPL}[1]{\small\textbf{Streszczenie:} #1}
\newcommand{\abstractEN}[1]{\small\textbf{Abstract:} #1}
\usepackage{xcolor}
\usepackage{color}

\usepackage{caption}
\usepackage{subcaption}

\begin{document}\thispagestyle{empty}\pagestyle{fancy}
\begin{minipage}[t]{0.5\textwidth}\vspace{0pt}%
\includegraphics[scale=1.1]{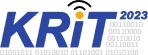}
\end{minipage}
\begin{minipage}[t]{0.45\textwidth}\vspace{0pt}%
\centering
KONFERENCJA RADIOKOMUNIKACJI\\ I TELEINFORMATYKI\\ KRiT 2023
\end{minipage}

\vspace{1cm}

\begin{center}
\titlePL{Badanie Sieci Massive MIMO o~Architekturze Zorientowanej na Użytkownika}

\titleEN{Evaluation of User-Centric Massive MIMO Network}\medskip

Marcin Hoffmann$^{1}$;
%Paweł Kryszkiewicz$^{2}$;

\medskip

\begin{minipage}[t]{0.6\textwidth}
\small $^{1}$ Politechnika Poznańska, Poznań, \href{mailto:email}{marcin.hoffmann@put.poznan.pl}\\
%\small $^{2}$ Politechnika Poznańska, Poznań, \href{mailto:email}{pawel.kryszkiewicz@put.poznan.pl}\\
\end{minipage}

\medskip

\end{center}

\medskip

\begin{multicols}{2}
\noindent
\abstractPL{
W nowoczesnych sieciach mobilnych 6G, wykorzystujących technikę M-MIMO (ang. Massive Multiple-Input-Multiple-Output) prawdopodobne jest zastosowanie architektury zorientowanej na użytkownika, w~której każdy użytkownik jest obsługiwany jednocześnie przez wszystkie stacje bazowe. W~tej pracy taki system został zbadany z użyciem zaawansowanego symulatora, wykorzystującego model kanału z tzw. trójwymiarowym śledzeniem promieni. Otrzymane wyniki pokazują, że architektura zorientowana na użytkownika zwiększa przepływności osiągane przez użytkowników charakteryzujących się najsłabszymi warunkami radiowymi ponad trzykrotnie.\footnote[1]{Praca powstała w~ramach projektu PRELUDIUM  finansowanego przez Narodowe Centrum Nauki nr 2022/45/N/ST7/01930.}}
\medskip

\noindent
\abstractEN{
The future 6G networks are expected to utilize large antenna arrays and follow the user-centric architecture, where the user is being served by all base stations. This work evaluates such a~system within an advanced system-level simulator, which utilizes an accurate 3D Ray-Tracing radio channel model. Results show that the novel user-centric network architecture can increase the cell-edge users' throughput by a~fold of 3.}
\medskip

\noindent
\keywordsPL{6G, Massive MIMO, sieć zorientowana na użytkownika, trójwymiarowe śledzenie promieni}
\medskip

\noindent
\keywordsEN{6G, Massive MIMO, User-Centric networks, 3D Ray-Tracing}

\section{Wstęp}

Żeby sprostać rosnącym wymaganiom użytkowników sieci mobilnych konieczne jest wykorzystanie technik transmisyjnych umożliwiających zwiększenie efektywności widmowej. Dla obecnie wdrażanych sieci 5G i~przyszłych sieci 6G taką techniką jest tzw. M-MIMO (\textit{ang. Massive Multiple-Input-Multiple-Output})~\cite{akyildiz2020}. Zakłada ona wykorzystanie wieloelementowych macierzy antenowych, które pozwalają na tworzenie wąskich wiązek sygnału radiowego skierowanych do poszczególnych użytkowników. Dzięki temu możliwa jest multipleksacja przestrzenna użytkowników, którym przydzielone są te same zasoby czasowo-częstotliwościowe. Systemy 5G wykorzystujące M-MIMO typowo zakładają obsługę jednego użytkownika przez jedną stacje bazową. Jest to tzw. architektura zorientowana na sieć. W~takiej architekturze pokrycie sygnałem radiowym jest nierównomierne: użytkownicy znajdujący się blisko stacji bazowej odbierają znacznie silniejszy poziom sygnału, od tych, którzy znajdują się dalej.
%maja jednak architekturę zorientowaną na sieć, tzn. występują granice pomiędzy komórkami, a~użytkownik typowo obsługiwany jest przez jedną stację bazową. 
W związku z nierównomiernym pokryciem komórek sygnałem radiowym, istnieje silna dysproporcja pomiędzy przepływnościami oferowanymi użytkownikom znajdującym się blisko stacji bazowej, a~tymi będącymi na skraju komórki. %Co więcej, użytkownicy znajdujący się na skraju komórki doświadczają silnej interferencji z sąsiednich komórek. 
Z tego względu dla systemów 6G rozważana jest architektura zorientowana na użytkownika, połączona z techniką M-MIMO (\textit{ang. User-Centric-Cell-Free M-MIMO})~\cite{cfmimobook2021}. W~sieci M-MIMO zorientowanej na użytkownika znika koncepcja komórek, a~użytkownik może być obsługiwany przez wszystkie otaczającego go stacje bazowe. W~ten sposób użytkownik znajduje się efektywnie w~środku wirtualnej komórki, co pozwala znacznie zmniejszyć dysproporcje pomiędzy przepływnościami różnych użytkowników. %osiągającymi najwyższe przepływności, a~tymi którzy znajdują się na skraju geograficznych komórek (pomiędzy kilkoma stacjami bazowymi). 
Wiele prac, w~których autorzy badają możliwości systemów o~architekturze zorientowanej na użytkownika zakłada bardzo proste modele systemu~\cite{ngo2015}, tzn. system jest wąskopasmowy, podczas gdy systemy 5G powszechnie wykorzystują modulacje wielotonowe i~wielodostęp z ortogonalnym podziałem częstotliwości OFDMA (\textit{ang. Orthogonal Frequency-Division Multiple Access}). Ponadto stosowane są proste, statystyczne modele kanałów radiowych często nieuwzględniające istotnych z punktu widzenia M-MIMO korelacji przestrzennych między użytkownikami. Istnieje więc zapotrzebowanie na zbadanie systemów M-MIMO o~architekturze zorientowanej na użytkownika w~środowisku możliwie bliskim rzeczywistej sieci, np. uwzględniających wykorzystanie techniki OFDMA oraz dokładny model kanału radiowego bazujący na tzw. trójwymiarowym śledzeniu promieni (\textit{ang. 3D Ray-Tracing}). Ponadto należy rozważyć wielowarstwowość przyszłych systemów 6G składających się z różnych bloków funkcyjnych odpowiedzialnych m.in. za przydział zasobów czasowo częstotliwościowych, prekodowanie i~dobór schematów modulacji i~kodowania. Celem tej pracy jest zbadanie zysków wynikających z zastosowania architektury sieci M-MIMO zorientowanej na użytkownika w~porównaniu do tradycyjnej architektury zorientowanej na sieć za pomocą zaawansowanego symulatora komputerowego, opracowanego przez autora.
W dalszych częściach tej pracy, w~Rozdziale \ref{sec:uccf_network} przedstawiona zostanie koncepcja sieci M-MIMO o~architekturze zorientowanej na użytkownika. W~Rozdziale \ref{sec:implementation_challenges} omówione będą wyzwania związane z praktyczną implementacją takiej architektury sieci. Rozdział \ref{sec:simulation_environment} zawiera opis zaproponowanego zaawansowanego środowiska symulacyjnego, służącego do badania sieci M-MIMO zorientowanej na użytkownika. Wyniki symulacji przedstawione są w~Rozdziale \ref{sec:results}, a~wnioski są sformułowane w~Rozdziale~\ref{sec:conclusions}.

\section{Sieć Zorientowana na Użytkownika} \label{sec:uccf_network}

\begin{figure}[H]
\centering
\includegraphics[scale=0.4]{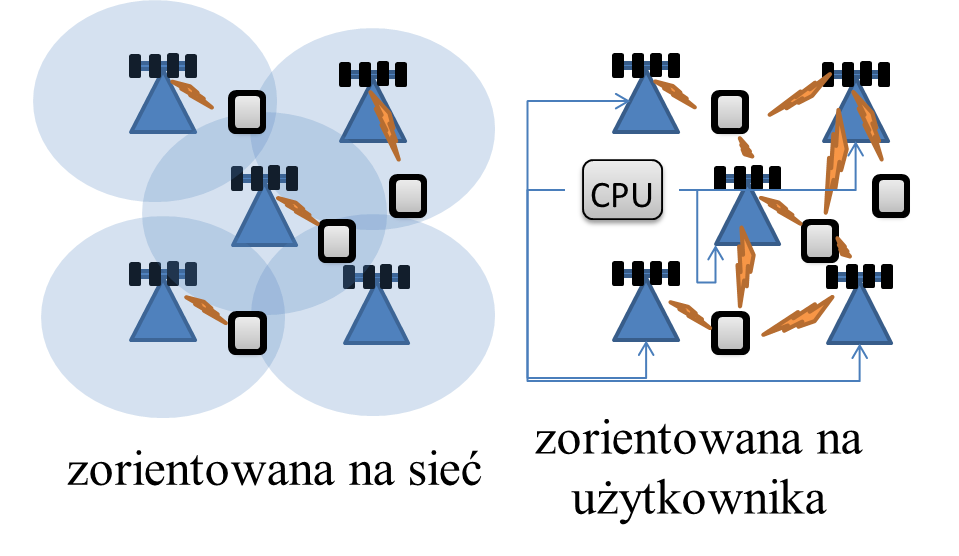}
\caption{Architektury systemu M-MIMO.}
\label{fig:architecture}
\end{figure}

Na Rys.~\ref{fig:architecture} architektura systemu M-MIMO zorientowanego na sieć została porównana z architekturą zorientowaną na użytkownika. W~pierwszym przypadku użytkownik obsługiwany jest przez jedną ze stacji bazowych. Typowo jest to stacja od której odbiera najsilniejszy sygnał referencyjny. W~tym przypadku przydział zasobów radiowych oraz całość przetwarzania sygnału (kodowanie kanałowe, prekodowanie, modulacja etc.) odbywa się niezależnie w~obrębie każdej komórki. Nawet jeśli specyfikacje techniczne~\cite{DAHLMAN201839} umożliwiają pewną współpracę pomiędzy komórkami, np. w~celu koordynacji interferencji, dotyczy ona tylko kilku stacji bazowych. Ponadto utworzony w~ten sposób klaster stacji bazowych obsługujących użytkownika nie może być dynamicznie modyfikowany, przez co pokrycie sygnałem radiowym obszarów znajdujących się na granicy klastra jest zdecydowanie mniejsze niż w~jego centrum. W~związku z tym w~systemach o~architekturze zorientowanej na sieć występuje silna dysproporcja pomiędzy przepływnościami osiąganymi przez użytkowników znajdujących się na skraju komórki i~tymi będącymi blisko jej centrum. Architektura zorientowana na użytkownika zakłada zgodnie z definicją podaną w~\cite{ngo2015}, że użytkownik jest obsługiwany jednocześnie przez wszystkie stacje bazowe. Co więcej, sygnał nadawany przez wiele stacji bazowych jest odbierany przez użytkownika w~sposób koherentny. Stacje bazowe podłączone są do tzw. Centralnej Jednostki Przetwarzania CPU (\textit{ang. Central Processing Unit}). W~zależności od implementacji podział przetwarzania sygnału pomiędzy stacje bazowe a~CPU może się różnić. W~przypadku pełnej centralizacji całość przetwarzania realizowana jest w~CPU, a~poszczególne stacje bazowe pełnią tylko funkcje tzw. zdalnych jednostek radiowych RRU (\textit{ang. Remote Radio Unit}). Rozproszone implementacje przesuwają cześć przetwarzania sygnału do stacji bazowych np. prekodowanie sygnału, modulacja, kodowanie kanałowe. Zaletą implementacji scentralizowanej jest możliwość holistycznej optymalizacji transmisji w~systemie, np. koordynacji interferencji. Wadą natomiast jest konieczność przesyłania dużej ilości danych kontrolnych i~informacji, np. o~współczynnikach kanałów radiowych.

Dla pojedynczego bloku zasobów o~indeksie $n$, moc sygnału odbieranego przez $k$-tego użytkownika w~systemie M-MIMO zorientowanym na użytkownika, składającym się z $N_{\mathrm{bs}}$ stacji bazowych (często nazywanych również punktami dostępu (\textit{ang. Access Points}) dana jest wzorem~\cite{cfmimobook2021}:
\begin{equation}\label{eq:wanted_power}
    P_{k,n} = \left| \sum_{l=1}^{N_{\mathrm{bs}}} \sqrt{p_{kln}} \boldsymbol{h}^H_{kln}\boldsymbol{w}_{kln}\right|^2,
\end{equation}
gdzie $p_{kln}$ jest mocą nadawaną w~pasmie $n$-tego bloku zasobów przez stację bazową $l$ do $k$-tego użytkownika, $\boldsymbol{h}_{kln}$ jest wektorem zespolonych współczynników kanału radiowego pomiędzy $k$-tym użytkownikiem, a~każdą z $M_l$ anten stacji bazowej $l$, stałych w~pasmie $n$-tego bloku zasobów, a~$\boldsymbol{w}_{kln}$ jest wektorem wag prekodera. Można zauważyć, że sygnał nadawany przez wszystkie stacje bazowe może się koherentnie dodawać podczas odbioru przez użytkownika, co ma szczególne znaczenie w~przypadku, gdy znajduje się on w~podobnej odległości od kilku stacji bazowych (na granicy geograficznej komórki, w~sensie systemu zorientowanego na sieć). Należy jednak zauważyć, że podobną własność ma całkowita moc interferencji generowanej przez innych użytkowników, która dana jest wzorem~\cite{cfmimobook2021}:
\begin{equation}\label{eq:interference_power}
    I_{k,n} = \sum_{i=1}^{K_n} \left(\left| \sum_{l=1}^{N_{\mathrm{bs}}} \sqrt{p_{iln}} \boldsymbol{h}^H_{kln}\boldsymbol{w}_{iln}\right|^2 \right)- P_{k,n},
\end{equation}
gdzie $K_n$ oznacza liczbę użytkowników, którym przydzielono $n$-ty blok zasobów. Jak już wspomniano interferencje można znacznie ograniczyć w~przypadku scentralizowanej implementacji systemu M-MIMO zorientowanego na użytkownika.
\section{Wyzwania Implementacyjne} \label{sec:implementation_challenges}

System M-MIMO o~architekturze zorientowanej na użytkownika pozwala uzyskać bardziej równomierny rozkład przepływności osiąganych przez użytkowników, kosztem szeregu wyzwań stojących przed praktyczną implementacją takiej architektury przyszłych sieci mobilnych. Do najważniejszych z nich nalezą~\cite{ammar2022}:
    \paragraph{Wirtualne Klastry Obsługi}
    W~praktyce często może okazywać się, że obsługa każdego użytkownika przez wszystkie stacje bazowe stwarza konieczność wymiany bardzo dużej ilości danych sygnalizacyjnych i~znacznie zwiększa złożoność obliczeniową przetwarzania sygnału, np. konieczne jest wymiana informacji dotyczących kanałów radiowych użytkowników. W~wielu przypadkach część stacji bazowych będzie miała niewielki wpływ na wzrost przepływności danego użytkownika, np. z powodu znacznej odległości od niego. Z tej perspektywy istotne jest tworzenie tzw. wirtualnych klastrów obsługi (\textit{ang. Serving Clusters}), czyli podzbiór wszystkich stacji bazowych, które obsługują danego użytkownika.   
    \paragraph{Opóźnienia i~synchronizacja} W~związku z tym, że poszczególne stacje bazowe jednocześnie obsługujące użytkowników są od siebie oddalone o~nawet kilka kilometrów istotnym problem mogą się okazać opóźnienia oraz synchronizacja. Problem dotyczy zarówno wymiany informacji sygnalizacyjnej pomiędzy stacjami bazowymi a~CPU, jak i~warstwy fizycznej, np. dobranie wag prekodera w~niektórych przypadkach będzie musiało uwzględniać brak synchronizacji fazy pomiędzy poszczególnymi stacjami bazowymi.  
    \paragraph{Alokacja Zasobów} W~systemach zorientowanych na użytkownika potrzebne jest inne podejście do przydziału zasobów radiowych niż w~przypadku systemów o~architekturze zorientowanej na sieć. Dla systemu zorientowanego na użytkownika alokacja zasobów odbywa się w~ramach całej sieci (lub wirtualnego klastra obsługi), a~nie pojedynczej komórki. Ważnym zagadnieniem dotyczącym alokacji zasobów jest dynamiczne dobieranie użytkowników w~grupy, które mogą podlegać multipleksacji przestrzennej (\textit{ang. User Pairing}). 
    \paragraph{Efektywność Energetyczna} Sieć zorientowana na użytkownika musi brać pod uwagę zużycie energii potrzebne do zapewnienia odpowiedniej jakości usług. Zarówno zasoby sprzętowe w~CPU, jak i~te zainstalowane na stacjach bazowych muszą się w~inteligentny sposób skalować i~dostosowywać do aktualnego ruchu w~sieci. Efektywność energetyczna może być zwiększana, np. poprzez tworzenie wirtualnych klastrów obsługi w~taki sposób, żeby niektóre stacje bazowe mogły być wprowadzane w~stan uśpienia. Szczególne znaczenie w~tym kontekście ma wykorzystanie metod uczenia maszynowego i~sztucznej inteligencji%, ze szczególnym naciskiem na tzw. uczenie ze wzmocnieniem RL (\textit{ang. Reinforcement Learning})
    ~\cite{Ghiasi2023}.
%    \item \textbf{Estymacja Kanału Radiowego}
%\paragraph{Uczenie Maszynowe}\\ Warto zaznaczyć, że istotna rolę w~implementacji i~zarządzaniu systemem M-MIMO zorientowanym na użytkownika będzie miało wykorzystanie metod uczenia maszynowego i~sztucznej inteligencji, ze szczególnym naciskiem na tzw. uczenie ze wzmocnieniem RL (\textit{ang. Reinforcement Learning})~\cite{Ghiasi2023}. Algorytmy RL bazują na interakcji tzw. agenta, ze środowiskiem i~mogą być z powodzeniem wykorzystywane do np. w~celu optymalizacji efektywności energetycznej bazującej na informacji o~poleżeniu użytkowników. 
\section{Opracowany Symulator} \label{sec:simulation_environment}

Na potrzeby porównania systemów M-MIMO o~architekturze zorientowanej na użytkownika, z tymi o~architekturze zorientowanej na sieć opracowany został zaawansowany symulator komputerowy. Jest to symulator systemowy (\textit{ang. system-level simulator}), który implementuje łącze w~dół (\textit{ang. Downlink}) w~systemie OFDM składającym się z 6 stacji bazowych podłączonych do CPU. System zajmuje pasmo o~szerokości $25$~MHz zaalokowane wokół częstotliwości środkowej równej $3.6$~GHz i~podzielone na $N_\mathrm{rb}=69$ bloków zasobów. W~systemie znajduje się jedna stacja bazowa typu makro (oznaczona indeksem $l=0$) zainstalowana na wysokości $45$~m, której całkowita moc nadawana jest równa $P^\mathrm{tx}_0=46$~dBm i~jest wyposażona w~macierz składającą się z $M_0=128$ anten ($8$ rzędów $\times$ 16 kolumn).%, zainstalowaną na wysokości $45$~m. 
Pozostałe 5 stacji bazowych to stacje typu mikro zainstalowane na wysokości $6$~m. Każda z nich nadaje z mocą $P^\mathrm{tx}_l=30$~dBm i~jest wyposażona w~dwie macierze antenowe o~rozmiarze $M_l=16$ (8 rzędów $\times$ 2 kolumny). %Stacje mikro są zainstalowane na wysokości $6$~m. 
Model ruchu wykorzystywany w~symulatorze to tzw. model pełnego bufora (\textit{ang. Full Buffer}), w~którym użytkownikom rozdzielane są wszystkie zasoby radiowe dostępne w~systemie. Przydział zasobów radiowych odbywa się w~formie scentralizowanej niezależnie dla każdego bloku zasobów, według procedury maksymalizującej tzw. metrykę PF (\textit{ang. Proportional Fairness})~\cite{yoo2006}. W~ramach każdego bloku zasobów użyty algorytm sprawdza poziom korelacji kanałów radiowych użytkowników i~na tej podstawie dokonuje multipleksacji w~dziedzinie przestrzennej. Moc nadawana przez każdą stację bazową do zaalokowanych użytkowników zależy od ich warunków propagacyjnych i~dana jest wzorem:
\begin{equation}
    p_{kln} = \frac{P^\mathrm{tx}_l}{N_\mathrm{rb}} \cdot  \frac{||\boldsymbol{h}_{kln}||^2}{\sum_{i=1}^{K_n} ||\boldsymbol{h}_{iln}||^2},
\end{equation}
gdzie $||\cdot||$ oznacza normę $l^2$ z wektora. W~ten sposób żadna stacja bazowa nie będzie nadawała dużej mocy do użytkowników znajdujących się daleko od niej. Wszystkie stacje bazowe wykorzystują prekoder MRT (\textit{ang. Maximum Ratio Transmission}), który dla każdego użytkownika maksymalizuje jego współczynnik sygnału do szumu:% SNR (\textit{Signal to Noise Ratio}):
\begin{equation}
    \boldsymbol{w}_{kln} = \frac{\boldsymbol{h}_{kln}}{||\boldsymbol{h}_{kln}||}.
\end{equation}
W symulatorze zaimplementowanych zostało 15 schematów modulacji i~kodowania MCS (\textit{ang. Modulation and Coding Schemes}), dobieranych według procedury opisanej w~\cite{bossy2017}, na podstawie stosunku sygnału do szumu i~interferencji (obserwowanych i~raportowanych przez użytkowników, oraz estymowanych na etapie przydziału zasobów radiowych) obliczonego z użyciem wzorów~\eqref{eq:wanted_power}~i~\eqref{eq:interference_power}.
Zaimplementowane środowisko symulacyjne wykorzystuje realistyczny model kanału radiowego bazujący na trójwymiarowym śledzeniu promieni dostarczony przez firmę Wireless InSite\textsuperscript{TM}. Kanał radiowy został wygenerowany w~scenariuszu miejskim zdefiniowanym w~ramach projektu METIS, i~jest to tzw. model \emph{Madrid Grid}~\cite{metis2020}. Każda ścieżka obliczona jest przy założeniu maksymalnie 15 odbić i jednej dyfrakcji.  

\section{Wyniki Symulacji} \label{sec:results}
\begin{figure}[H]
\centering
\begin{subfigure}[b]{0.501\textwidth}
\centering
\includegraphics[scale=0.18]{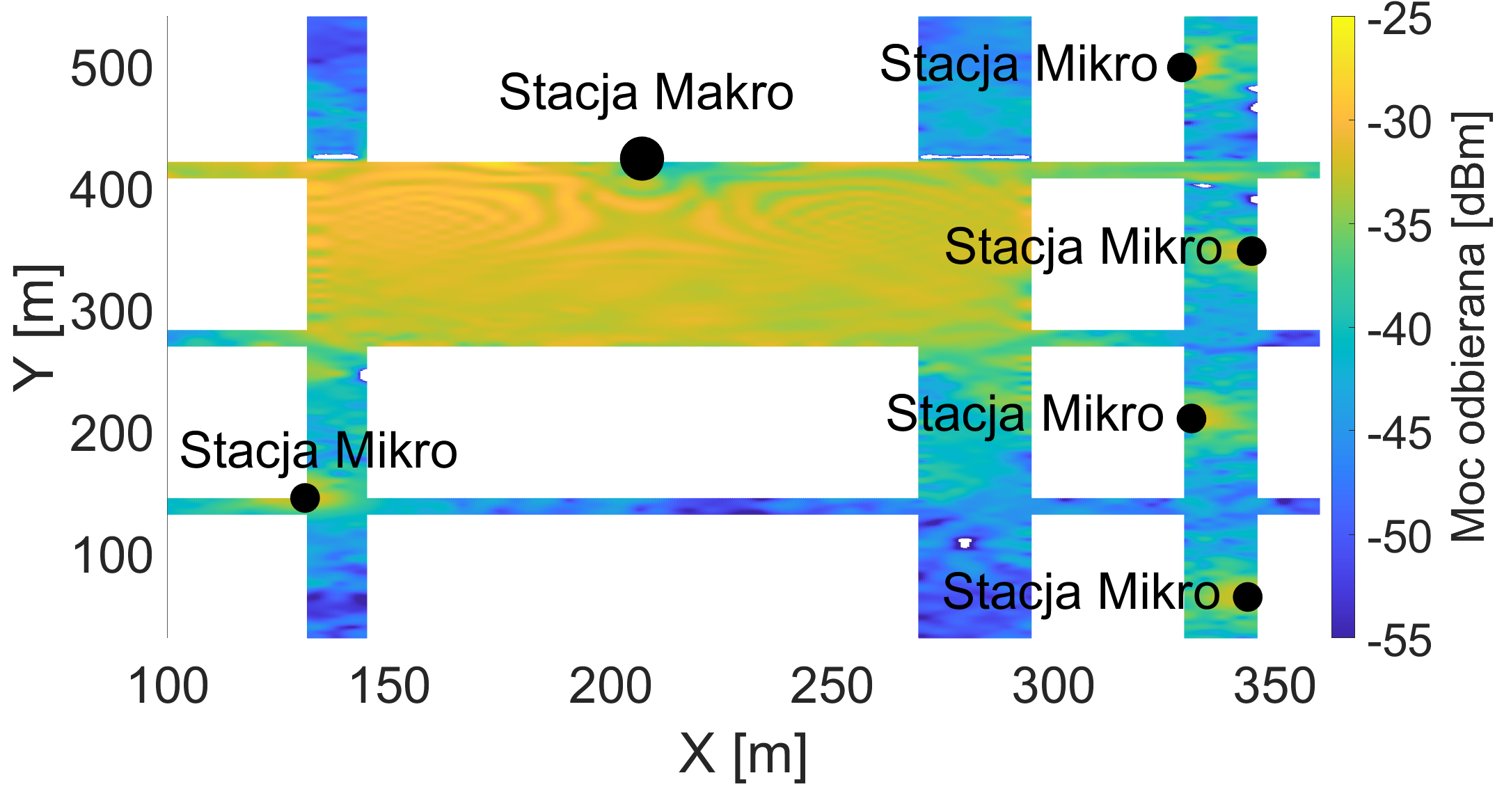}
\caption{Systemu o~architekturze zorientowanej na sieć}
\label{fig:map_1ap}
\end{subfigure}%

\begin{subfigure}[b]{0.501\textwidth}
\centering
\includegraphics[scale=0.18]{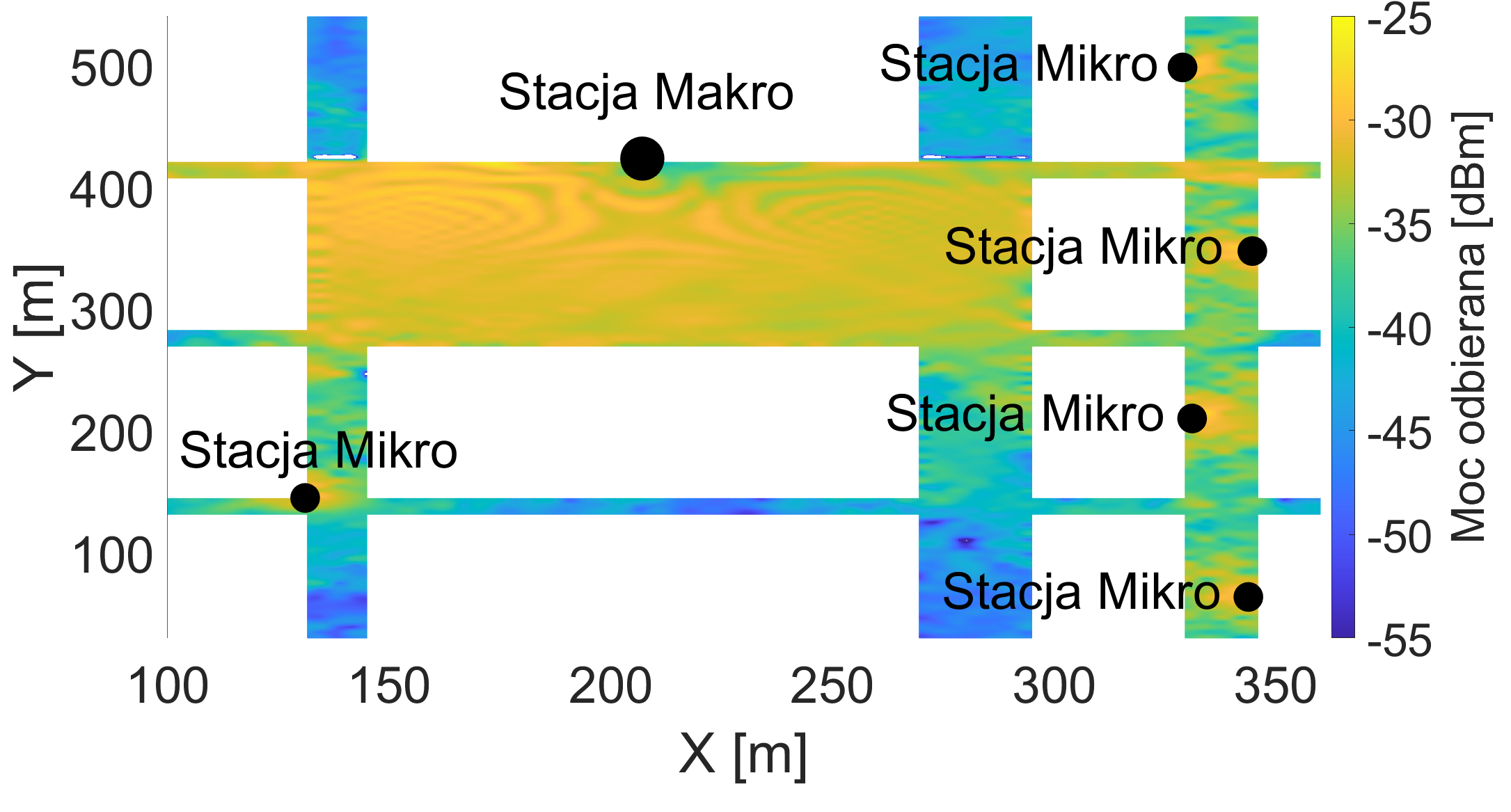}
\caption{System o~architekturze zorientowanej na użytkownika }
\label{fig:map_uccf}
\end{subfigure}
\caption{Maksymalna moc sygnału odbieranego}%\label{fig:animals}
\end{figure}
\begin{figure}[H]
\centering
\includegraphics[width=3.1in]{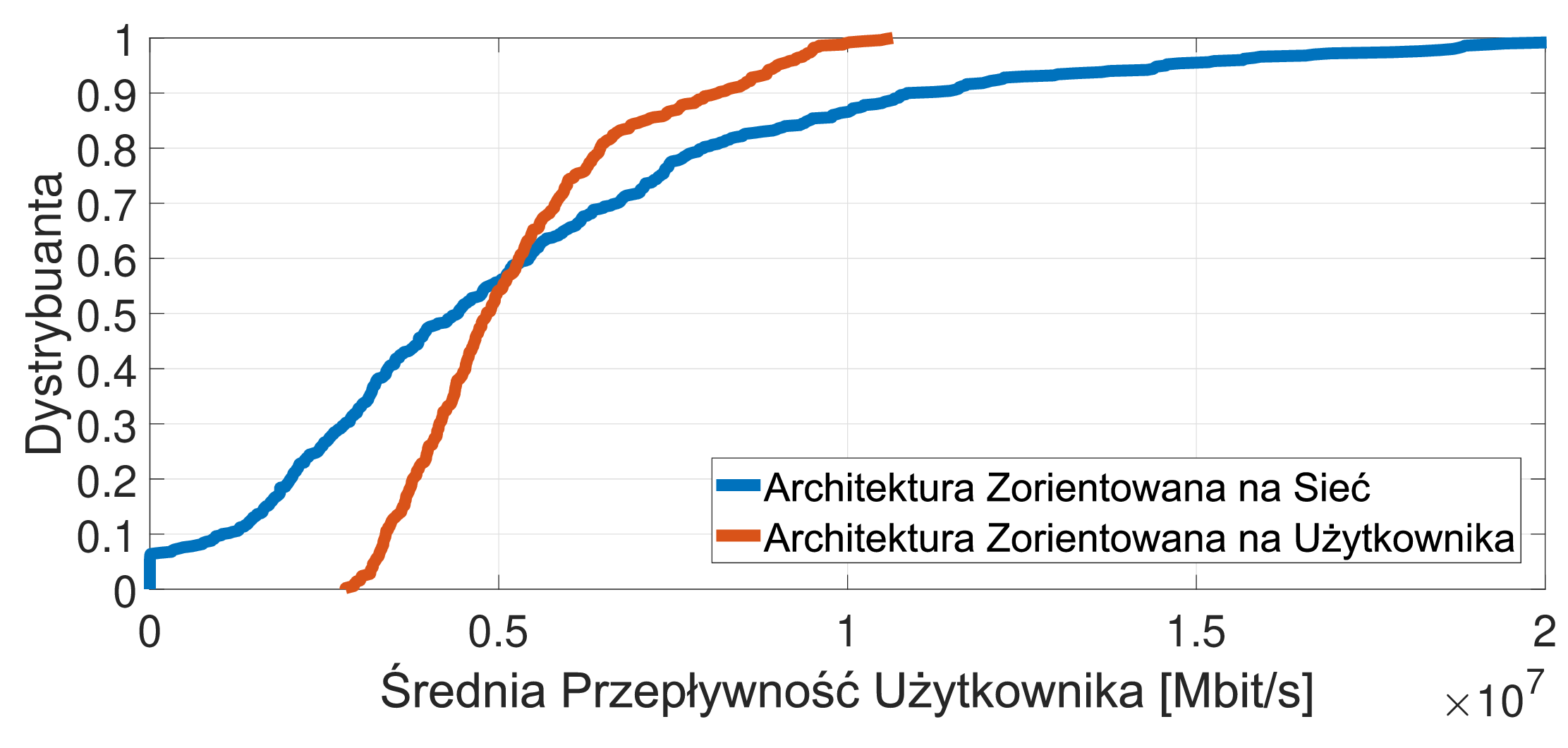}
\caption{Porównanie dystrybuant rozkładu średnich przepływności użytkowników, dla systemu o~architekturze zorientowanej na użytkownika i~na sieć}
\label{fig:cdf_rate}
\end{figure}
\begin{figure}[H]
\centering
\includegraphics[width=3.1in]{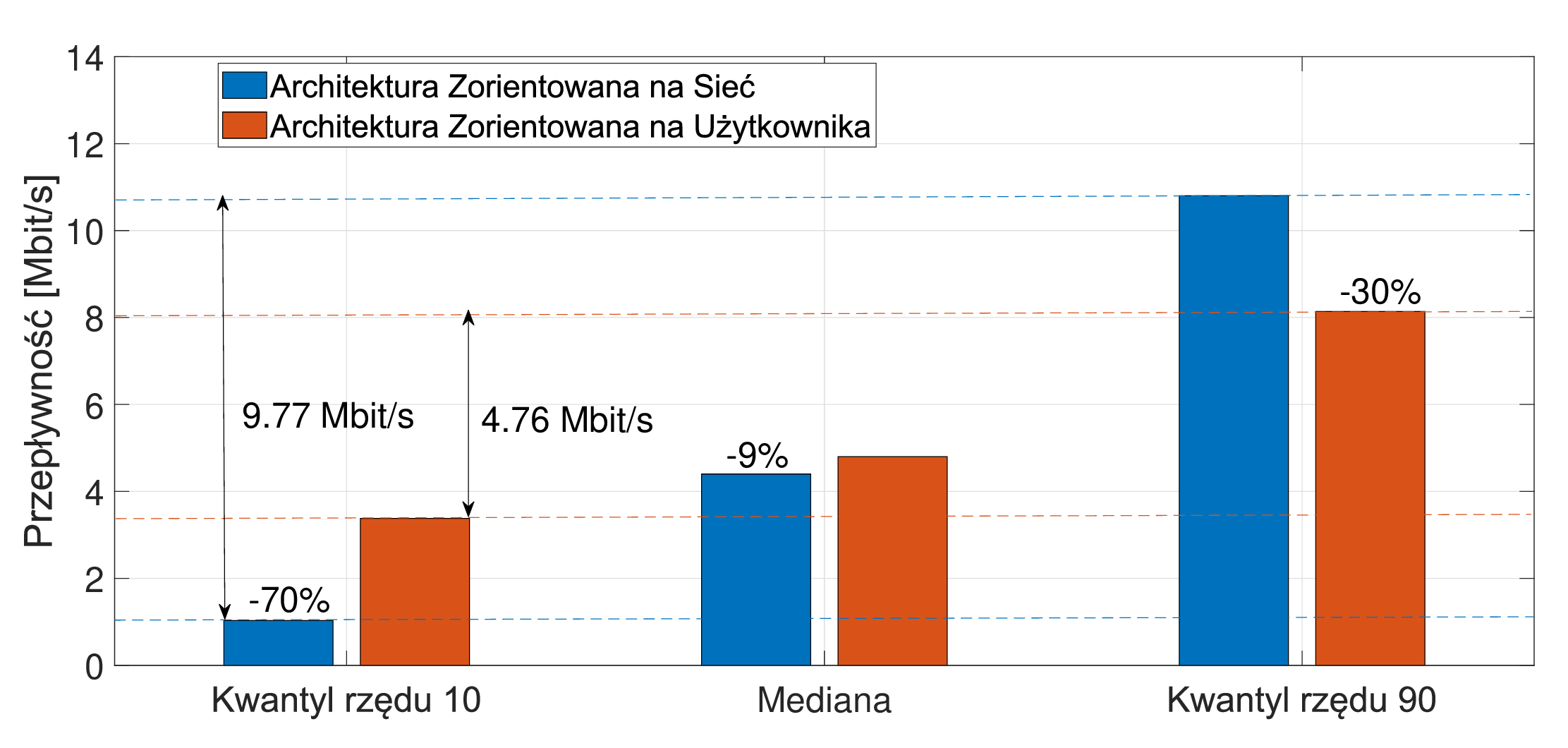}
\caption{Porównanie kwantyli 10,50 i~90 rozkładu średnich przepływności użytkowników, dla systemu o~architekturze zorientowanej na użytkownika i~na sieć}
\label{fig:bar_rate}
\end{figure}

W ramach eksperymentu symulacyjnego,
najpierw porównane zostało pokrycie sygnałem radiowym rozważnego obszaru w~przypadku architektury zorientowanej na sieć i~na użytkownika. W~tym celu wygenerowany został kanał radiowy dla użytkowników rozmieszczonych równomiernie na obszarze sieci w~odstępach co $5$~m (łącznie 3542 użytkowników). Dla każdego użytkownika z użyciem wzoru~\eqref{eq:wanted_power} i przy założeniu prekodera MRT obliczony został poziom odbieranej mocy sygnału w~przypadku gdy obsługuje go jedna stacja bazowa, oraz w~przypadku sieci zorientowanej na użytkownika, kiedy obsługiwany jest przez wszystkie stacje bazowe. Wyniki uśrednione po wszystkich blokach zasobów zostały przedstawione odpowiednio na Rys.~\ref{fig:map_1ap} i Rys.~\ref{fig:map_uccf}, gdzie biały kolor oznacza budynki. Zalety zastosowania architektury zorientowanej na użytkownika są szczególnie widoczne w~alejce z prawej strony, pokrytej przez liczne stacje mikro. Dla architektury zorientowanej na użytkownika alejka jest pokryta równomiernie sygnałem radiowym o~mocy oscylującej wokół -35~dBm. W~przypadku architektury zorientowanej na sieć widoczne są wyraźne granice pomiędzy komórkami, gdzie moc odbierana spada poniżej -50~dBm. Dzięki zastosowaniu architektury zorientowanej na użytkownika na granicach komórek użytkownicy mogą odbierać moc wyższą o~nawet 15 dB niż w~przypadku architektury zorientowanej na sieć.
Następnie przeprowadzone zostały symulacje systemowe z użyciem środowiska opisanego w~Rozdziale~\ref{sec:simulation_environment}. Przeprowadzono 10 przebiegów symulacyjnych odpowiadających 0.5~s działania rzeczywistego systemu 5G (1000 slotów trwających 0.5~ms każdy). W~ramach każdego przebiegu symulacyjnego rozważonych jest 50 użytkowników losowo rozmieszczonych na obszarze sieci i~poruszających się w~losowych kierunkach z prędkością 1.5~m/s. Na koniec każdego przebiegu symulacyjnego zebrane zostały statystyki dotyczące średniej przepływności każdego z rozważanych użytkowników. Dane o~średniej przepływności użytkowników zebrane podczas 10 przebiegów symulacyjnych posłużyły do stworzenia rozkładu średnich przepływności użytkowników. Eksperyment symulacyjny w~tych samych warunkach (dla tych samych jąder generatorów liczb losowych) przeprowadzono rozważając architekturę zorientowaną na użytkownika i~na sieć.  Na Rys.~\ref{fig:cdf_rate} porównane zostały dystrybuanty rozkładu średniej przepływności użytkowników w~obu rozważanych architekturach systemu M-MIMO. Widać wyraźnie, że w~przypadku architektury systemu zorientowanej na użytkownika przepływności osiągane przez użytkowników są bardziej równomierne niż w~przypadku architektury zorientowanej na sieć. Ponadto widać dużą poprawę przepływności dla użytkowników mających najbardziej wymagające warunki radiowe. Na Rys.~\ref{fig:bar_rate} porównane zostały kwantyle rzędu 10, 50 (mediana) i~90 z rozkładu średnich przepływności użytkowników. Architektura zorientowana na użytkownika pozwala poprawić medianę i~kwantyl rzędu 10 odpowiednio o~9\% i~ponad trzykrotnie względem architektury zorientowanej na sieć. Warto zauważyć, że w~przypadku architektury systemu M-MIMO zorientowanej na sieć występują duże dysproporcje pomiędzy kwantylem rzędu 10 a~90, tj. 9.77~Mbit/s, dzięki zastosowaniu architektury zorientowanej na użytkownika można zmniejszyć te dysproporcje o~połowę do poziomu 4.76~Mbit/s. Kosztem wynikającym z zastosowania architektury zorientowanej na użytkownika jest pogorszenie przepływności osiąganych przez 10\% użytkowników charakteryzujących się najlepszymi warunkami radiowymi (kwantyl rzędu 90) o~30\%. 
\section{wnioski} \label{sec:conclusions}
Nowoczesna architektura systemu M-MIMO pozwala jednocześnie obsługiwać użytkownika przez wszystkie stacje bazowe w~systemie. Dzięki temu możliwe jest bardziej równomierne pokrycie sygnałem sieci, które przekłada się na znaczną poprawę przepływności osiąganych przez użytkowników charakteryzujących się najmniej korzystnymi warunkami radiowymi. Wyniki symulacyjne pokazały, że zysk może być ponad trzykrotny.

\bibliographystyle{krit}
\bibliography{references}

\end{multicols}
\end{document}